\def\eg{{e.g., }}

\def\spose#1{\hbox to 0pt{#1\hss}}
\def\lta{\mathrel{\spose{\lower 3pt\hbox{$\mathchar"218$}}
     \raise 2.0pt\hbox{$\mathchar"13C$}}}
\def\gta{\mathrel{\spose{\lower 3pt\hbox{$\mathchar"218$}}
     \raise 2.0pt\hbox{$\mathchar"13E$}}}
\def\ge{\mathrel{\spose{\lower 3pt\hbox{$-$}}
     \raise 2.0pt\hbox{$\mathchar"13E$}}}
\def\le{\mathrel{\spose{\lower 3pt\hbox{$-$}}
     \raise 2.0pt\hbox{$\mathchar"13C$}}}

\documentclass[10pt,preprint]{aastex}
\usepackage{graphics}
\usepackage{amsmath}
\usepackage{emulateapj5}

\newcommand {\acbar}{{\sc Acbar}}
\newcommand {\acbarsp}{{\sc Acbar} }

\slugcomment{Submitted to ApJ}

\shorttitle{{\sc Acbar} Parameter Estimates}
\shortauthors{Goldstein et al.}

\begin{document}

\title{ Estimates of Cosmological Parameters Using the CMB
 Angular Power Spectrum of ACBAR}

\author{J. H. Goldstein\altaffilmark{1,2},
P. A. R. Ade\altaffilmark{3},
J. J. Bock\altaffilmark{4},
J. R. Bond\altaffilmark{5},
C. Cantalupo\altaffilmark{6},
C. R. Contaldi\altaffilmark{5},
M. D. Daub\altaffilmark{7},
W. L. Holzapfel\altaffilmark{7},
C. Kuo\altaffilmark{7},
A. E. Lange\altaffilmark{8},
M. Lueker\altaffilmark{7},
M. Newcomb\altaffilmark{7},
J. B. Peterson\altaffilmark{9},
D. Pogosyan\altaffilmark{10},
J. E. Ruhl\altaffilmark{1}, 
M. C. Runyan\altaffilmark{8}, 
E. Torbet\altaffilmark{2}}
\altaffiltext{1}{Dept. of Physics, Case Western Reserve Univ., Cleveland, OH}
\altaffiltext{2}{Dept. of Physics, Univ. of California, Santa Barbara, CA}
\altaffiltext{3}{Cardiff University, UK}
\altaffiltext{4}{Jet Propulsion Laboratory, Pasadena, CA}
\altaffiltext{5}{Canadian Institute for Theoretical Astrophysics, University of Toronto, Canada}
\altaffiltext{6}{Lawrence Berkeley National Laboratory, Berkeley, CA 94720}
\altaffiltext{7}{University of California, 426 LeConte Hall, Berkeley, CA 94720-7300}
\altaffiltext{8}{California Institute of Technology, Pasadena, CA}
\altaffiltext{9}{Dept. of Physics, Carnegie Mellon University, 5000 Forbes Ave, Pittsburgh, PA 15213}
\altaffiltext{10}{Dept. of Physics, University of Alberta, Canada}

\begin{abstract}
We report an investigation of cosmological parameters based on the 
measurements of anisotropy in the cosmic microwave background
radiation (CMB) made by \acbar.  We use the \acbarsp data in concert
with other recent CMB measurements 
to derive Bayesian estimates of parameters in 
inflation-motivated adiabatic cold dark matter models. 
We apply a series of additional cosmological
constraints on the shape and amplitude of the density power spectrum,
the Hubble parameter and from supernovae to further refine our parameter estimates.  
Previous estimates of parameters
are confirmed, with sensitive measurements of the power
spectrum now ranging from $\ell \sim 3$ to $2800$.
Comparing individual best model fits, we find that the addition 
of $\Omega_{\Lambda}$ as a parameter dramatically improves the fits.
We also use the
high-$\ell$ data of \acbar, along with similar data from CBI and 
BIMA, to investigate potential secondary anisotropies from 
the Sunyaev-Zeldovich effect.  We show that the results from the three
experiments are consistent under this interpretation, and use the
data, combined and individually, to estimate $\sigma_8$ from the
Sunyaev-Zeldovich component. 

\end{abstract}


\section{Introduction}

Anisotropies in the cosmic microwave background radiation (CMB) are caused
by density and temperature fluctuations in the early universe, 
when radiation decoupled from matter ($z \sim 1100$). 
For models in which these initial fluctuations are of 
a Gaussian random nature, the information 
carried by the CMB anisotropies 
is completely characterized by their angular power spectrum as a
function of Legendre polynomial index $\ell$.  The comparison between
measured CMB angular power spectra and theoretical predictions 
can be used to rule out entire classes of cosmological models
as well as to estimate the values of a variety of cosmological parameters
within a given family of models.  

Acoustic oscillations that occurred before the universe became largely neutral
($z > 1100$) produced a harmonic series of broad peaks in the power
spectrum.  For flat models, the first peak lies near 
$\ell \sim 220$.  Superposed upon the subsequent peak/dip structure is an overall 
diminishment at higher $\ell$ because of photon
diffusion and the finite thickness of the last scattering surface. Above
$\ell \sim 1500$, the entire power spectrum is expected to be strongly 
attenuated by these effects;  this decay at high $\ell$ is 
called the damping tail. 

CMB anisotropies have been measured over a wide range of $\ell$ with
high significance.  Large angular scale observations
with the COBE-DMR instrument produced the first convincing detection of CMB
anisotropy~\citep{bennett92}.  At intermediate angular 
scales, a variety of groups
have made high signal-to-noise measurements of the first peak,
near $\ell \sim 200$~\citep{debernardis00,hanany00,halverson01,scott02}.
There is also very good evidence for additional harmonic features at 
higher $\ell$~\citep{halverson01,ruhl02}.  
Finally, at high $\ell$ the expected damping tail 
has been found~\citep{pearson02}, along with some
evidence for greater power at $\ell > 2000$ than is
expected for primary CMB anisotropies~\citep{mason02}.

These measurements provide strong support for the inflation-motivated
family of cold dark matter models with adiabatic initial density 
perturbations.  Improved measurements of the power spectrum will
put these models to a more stringent test.  The \acbarsp results 
reported in \citet[hereafter Paper~I]{kuo02}  provide the 
strongest constraint to date on the
angular power spectrum from $1000 < \ell < 2800$, in the damping tail
region.  In this paper we combine these new \acbarsp results with other 
published CMB results in order to investigate constraints on cosmological 
parameters in the adiabatic $\Lambda$-CDM model space.

The \acbarsp results provide detections of power in the
high-$\ell$ region where the effects of secondary anisotropies become
relevant. In this paper we also combine our high-$\ell$ data with previous results from the
CBI and BIMA interferometers operating at 30~GHz in an attempt to
quantify the contribution from the Sunyaev--Zeldovich Effect (SZE) to
the power spectrum.  Observations such as these, at multiple 
frequencies and 
over a range of angular scales, are essential to
the separation of contributions from the primary anisotropies and 
the SZE.

\section{The Instrument}
A brief description of the \acbarsp instrument is given 
in Paper~I, with a more complete treatment in~\cite{runyan02}.
We give only the most relevant details here.
\acbarsp is a 16 pixel bolometer array installed on the 2.1\,m Viper 
telescope at the South Pole, with an angular resolution of 
$\sim 5^{\prime}$ at 150~GHz.  We report here on results 
derived from a subset 
of those detectors, with bands centered at 150~GHz.  
In the first season of observations (2001) the 
array had four such 150~GHz detectors, while in the second season
(2002) there were eight.  Improvements to the receiver and
telescope between seasons led to greater sensitivity and 
improved pointing reconstruction in the second season.
  
The \acbarsp power spectrum used in this work is derived from 
observations of two fields covering a total of $24\,{\rm deg}^2$ of sky. 
The fields are selected to contain a bright guiding quasar near  
the center, which is removed prior to power spectrum estimation.
The absolute calibration of the instrument is derived from 
observations of Venus and Mars, and has an
uncertainty of 10\%.  
The beam profiles are derived from images of the guiding quasars 
and thus include any drifts or uncertainties in pointing reconstruction. 
The systematic uncertainty in the beam width, also needed here for use in
parameter estimation, is 3\%.

\section{The Power Spectrum}
\label{pssec}

\begin{figure*}[ht!]
\resizebox{\hsize}{!}{
\includegraphics{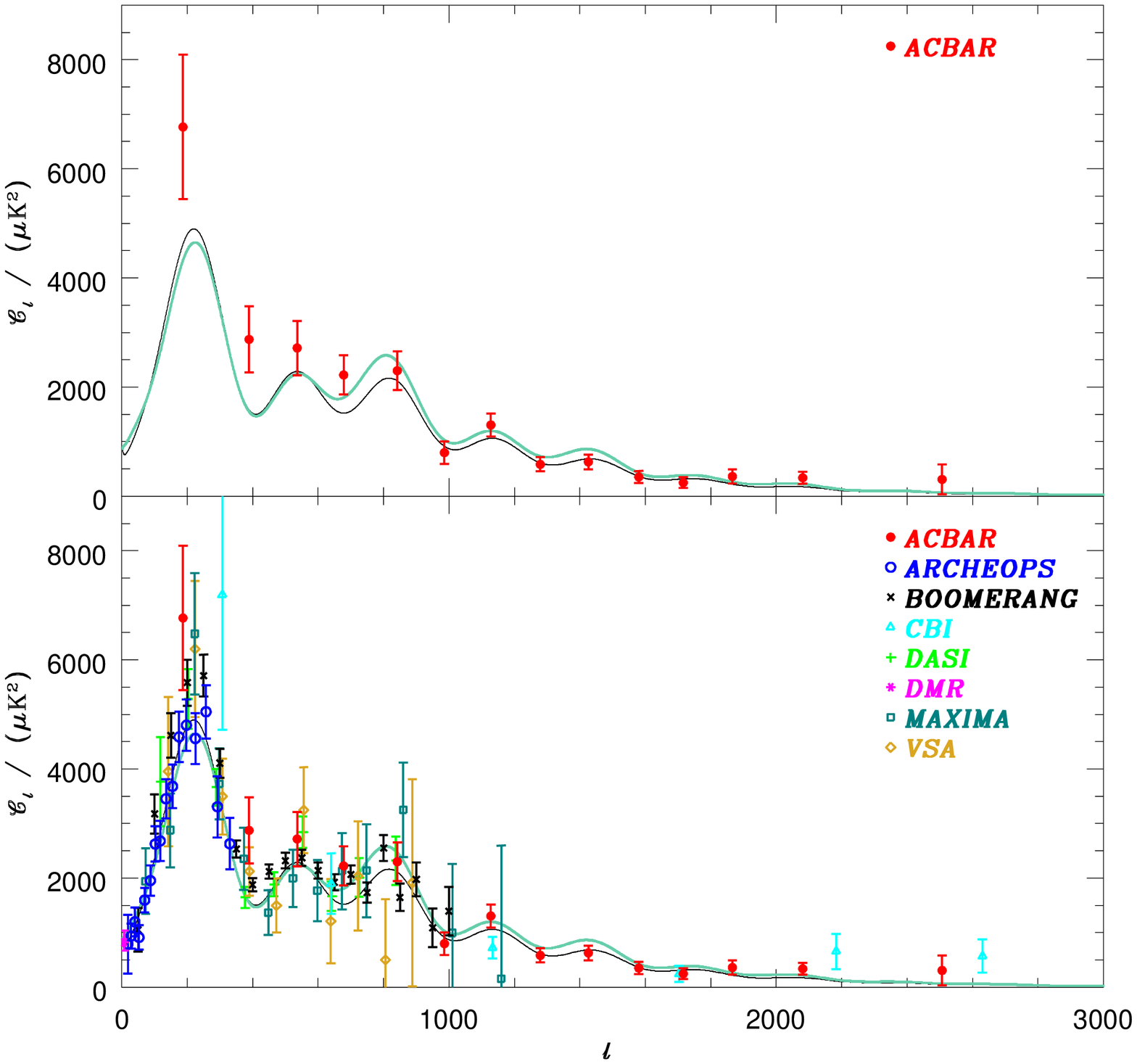}}
\caption{\protect\small
Top Panel:  The \acbar \ CMB power spectrum, ${\mathcal C}_{\ell} \equiv \ell
(\ell + 1)C_{\ell}/{(2\pi)}$, plotted over a vacuum energy
dominated [$\Omega_k = -0.05$, $\Omega_\Lambda = 0.5$, $\omega_{cdm} = 0.12$,
$\omega_b = 0.02$, $H_0 = 50$, $\tau_C = 0.025$, $n_s = 0.925$, amplitude
${\sqrt {\mathcal C}_{10}}= 1.11\times 10^{-5}T_{\rm CMB}$]
model (black thin line)
and a CDM dominated [$\Omega_k = 0.05$, $\Omega_\Lambda = 0$, 
$\omega_{cdm} = 0.22$, $\omega_b = 0.02$, $H_0 = 50$, $\tau_C = 0$,
$n_s = 0.925$, amplitude 
${\sqrt {\mathcal C}_{10}}= 1.34\times 10^{-5}T_{\rm CMB}$] model
(green thick line).
These are the best-fit models,
for $\Lambda$ and $\Lambda$-free models respectively,
found during the \acbar+Others parameter estimation described in the text,
with the weak-$h$ prior.
Bottom Panel: The top panel with the addition of power
spectra from several other experiments.  
Both models appear to be reasonable fits to the data, with the 
$\Omega_\Lambda = 0.5$ model statistically being the better of the two.}
\label{fig:aps}  
\end{figure*}

The results of Paper~I provide 14 ``band powers"
with effective centers between $\ell = 190$ and 2500.  
Figure~\ref{fig:aps} shows these results, along with a selection 
of other current measurements for context and comparison.
The points plotted in Figure~\ref{fig:aps} lie at the maxima $\overline{{\mathcal C}}_B$
of the joint likelihood distribution ${\mathcal L}(\{ {\mathcal C}_B)$
for each band power
(${\mathcal C}_B$), with the vertical bars
showing errors derived from the curvature of the 
likelihood at that maximum,
\begin{equation}
{\mathcal F}_{BB^{\prime}}^{({\mathcal C})} = 
	-\frac{\partial^2 \ln{{\mathcal L}({\mathcal C}_B)}}
	{\partial {\mathcal C}_B \partial {\mathcal C}_{B^{\prime}}} \ .
\end{equation}
The errors shown are Gaussian ones, 
$[{\mathcal F}^{-1/2}]_{BB}$. For \acbar, the curvature matrix
determined from the original bands has been rotated into the diagonal
frame, with eigenvalues $f_B$. Figure~\ref{fig:aps} shows $f_B^{-1/2}$
for the errors. For the other experiments shown, the results are
not rotated to the diagonal frame.

Of course the likelihood curves for the individual bandpowers are not
Gaussians.  An offset lognormal distribution~\citep{bond00} has been
shown to be an accurate representation of the likelihood curves, and
is used in all of our parameter estimations. This distribution is
Gaussian in the variables 
\begin{equation}\label{lognorm}
Z_B = \ln{({\mathcal C}_B + x_B)} \ . 
\end{equation}
Here, $x_B$ represents a noise contribution to the band power. For
\acbar, $x_B$ was treated as a parameter determined by fitting the
lognormal distribution to the likelihood, as described in in
Paper~I. $\overline{{\mathcal C}}_B$, $f_B^{-1/2}$, and $x_B$ are
given in Table~2 of that paper.

\section{Comparison with Theory}

For the standard cosmological model with adiabatic initial density
perturbations, the CMB angular power spectrum can be readily calculated
as a function of input cosmological parameters $\vec{y}$.
These theoretical predictions for $C_\ell$ can be made individually
for each $\ell$, while our measurement is over bands of finite width, 
characterized by the window functions $\varphi_{B\ell}$.  Given the
theoretically predicted $C_\ell$, the predicted 
band power ${\mathcal C}_B$ in one of our bands is
\begin{equation}\label{bandpow}
{\mathcal C}_B \equiv {\mathcal I}(\varphi_{B\ell}{\mathcal C}_{\ell})/
  {\mathcal I}(\varphi_{B\ell}) \ ,
\end{equation}
where ${\mathcal C}_{\ell} \equiv \ell(\ell + 1)C_{\ell}/2\pi$ and
\begin{equation}
{\mathcal I}(f_{\ell}) \equiv \sum\limits_{\ell}
  \frac{(\ell + \frac{1}{2})}{\ell(\ell + 1)}f_{\ell} \ .
\end{equation}

The window functions $\varphi_{B\ell}$ give the response of the bands
to power at each $\ell$.  Numerical tabulations of the window
functions are available on the \acbarsp experiment public
website\footnote{\tt
http://cosmology.berkeley.edu/group/swlh/acbar}. They have been
rotated into the diagonal frame.

The likelihood that the data would result from the cosmology described
by $\vec{y}$ is given by
\begin{equation}\label{entropy}\begin{split}
\ln{\mathcal L}({\bf{\mathcal C}}) =
&\ln{\mathcal L}(\overline{{\bf\mathcal C}}) -\\
&\frac{1}{2}\sum\limits_{BB^{\prime}}(Z_B - \overline{Z}_B)
{\mathcal F}_{BB^{\prime}}^{(z)}(Z_{B^{\prime}} -
\overline{Z}_{B^{\prime}}) \ ,
\end{split}\end{equation}
where $\overline{Z}_B =\ln{(\overline{{\mathcal C}}_B + x_B)}$ is the 
value of the lognormal parameter at the position of maximum likelihood
$\overline{{\mathcal C}}_B$ and ${\mathcal F}_{BB^{\prime}}^{(z)}$
is the curvature matrix transformed into the lognormal variables,
\begin{equation}
{\mathcal F}_{BB^{\prime}}^{(z)} = (\overline{{\mathcal C}}_B + x_B)
{\mathcal F}_{BB^{\prime}}^{({\mathcal C})}
(\overline{{\mathcal C}}_B^{\prime} + x_B^{\prime}) \ .
\end{equation}
Provided with the maximum likelihood band powers $\overline{{\mathcal C}}_B$,
the lognormal offsets $x_B$, the curvature matrix
${\mathcal F}_{BB^{\prime}}^{({\mathcal C})}$, and 
the window functions $\varphi_{B\ell}$, the likelihood of the
parameter set $\vec{y}$, given the data, can be computed.

Our set $\vec{y}$ consists of seven cosmological parameters:
{$\Omega_k=(1-\Omega_{tot}), \Omega_{\Lambda}, 
\omega_{cdm}, \omega_b, n_s, \tau_C,
\ln{{\mathcal C}_{10}}$}.  The total energy density of
the universe in units of critical density $\Omega_{tot}$ is linked to the 
global curvature of space:  negatively curved for
$\Omega_{tot} < 1$, positively curved for $\Omega_{tot} > 1$ and
flat for $\Omega_{tot} = 1$.  The total energy
density has 3 constituents: vacuum ($\Omega_\Lambda$), 
matter and relativistic particles.  The relativistic energy density is
currently negligible.  The matter
density is split into two types, baryonic matter ($\Omega_b \equiv 
\omega_b/h^2$), which interacts with electromagnetic radiation, and
cold dark matter ($\Omega_{cdm} \equiv \omega_{cdm}/h^2$),  which does not.
The total matter density is denoted $\Omega_m = \Omega_b + \Omega_{cdm}$.
The amplitude of the CMB power spectrum at
$\ell = 10$, $\ln{{\mathcal C}_{10}}$
gives the overall amplitude of the primordial fluctuations.
This quantity is well-constrained by the COBE-DMR
observations~\citep{bennett96}. The full COBE-DMR power spectrum as
described in~\citet{bond00}  
is included in all of our parameter analyses.
The spectral index of primordial density perturbations, $n_s$, parameterizes the
variation in the fluctuation power as a function of length scale; 
$n_s = 1$ corresponds to scale invariance. 

The universe reionized at some point between decoupling and the
present.  After reionization, CMB photons scatter further. $\tau_C$ is
the Compton optical depth (from decoupling to present) due to such
scattering.  High $\tau_C$ diminishes CMB power by a factor of
$\exp[-2\tau_C]$ over most of the $\ell$ range, though not in the DMR
range.   

 Many more parameters than our basic 7 may be needed to completely
describe inflationary models. These include the gravity-wave induced
tensor amplitude and tilt, variations of tilt with wavenumber,
relativistic particle densities, more complex dynamics associated with
the dark energy $\Omega_\Lambda$, etc.  For example, a gravity-wave
induced component is expected in the largest class of inflation
models, and has impact on the spectrum in the $\ell < 100$
region. Although far from the region where \acbarsp has its impact, it
can affect the amplitude, tilt, and Compton depth. If one constrains
the tilts and amplitudes of the tensor component to those motivated by
simple inflation models, there is little change in the other
parameters. 

It is possible that secondary anisotropies, such as the Sunyaev-Zeldovich
effect investigated later in this paper, could contribute significant 
power to the highest $\ell$ band of our measurement.   However, 
the detection of anisotropy in that band is only $1.1\sigma$, or
$0.9\sigma$ above the best-fit model primary CMB angular
power spectrum.  Thus, for the 
purpose of cosmological parameter estimation from the primary CMB signal,
we can safely ignore the effects of potential SZE contamination.

To derive estimates of cosmological parameters, we compare
our data with the primary CMB power spectra predicted
by combinations of cosmological parameters $\vec{y}$.   
In this comparison, we vary $\ln{{\mathcal C}_{10}}$ continuously, 
while the other parameters take the discrete values listed in 
Table~\ref{paramlist}.  
An angular power spectrum is generated for each set 
of discrete parameter values, forming a grid.  
Given a CMB dataset (\eg \acbarsp or a combination of various
measured power spectra), the likelihood is then calculated for 
each point on the seven dimensional grid.

To compute the likelihood that a particular parameter $X$ has 
a value $x_0$, the seven dimensional grid of likelihoods is 
integrated over the
other six parameters, holding the parameter of interest fixed at $x_0$.  
This method, known as marginalization,
involves calculating
\begin{equation}\label{prior}
L(X=x_0) = 
\int \delta(X-x_0)P_{prior}(\vec{y}){\mathcal L}(\vec{y})d\vec{y} \ ,
\end{equation}
where $\delta(x)$ is the usual delta function and the prior
$P_{prior}(\vec{y})$ is discussed below.

For each model on the parameter grid, along with the overall amplitude
parameter ${\cal C}_{10}$, we continuously vary the beam widths
$\sigma_{bi}$ and calibrations $A_i$ for each experiment $i$ about
their estimated values $\overline{\sigma_{bi}}$ and unity, to take into
account the uncertainties in the respective measurements.  We
approximate the beam-uncertainty and calibration-uncertainty
``prior'' probabilities by Gaussians in $\Delta (\sigma_{bi})^2$ and $\ln
A_i^2$, respectively. The modification to the bandpower as a result of
the uncertainty $\Delta (\sigma_{bi})^2$ is modeled by
$\exp[-\langle(\ell+\frac{1}{2})^2\rangle_B \Delta (\sigma_{bi})^2]$, with
$\langle(\ell+\frac{1}{2})^2\rangle_B =
{\mathcal I}(\varphi_{B\ell}(\ell+\frac{1}{2})^2)/
{\mathcal I}(\varphi_{B\ell}) $. The overall impact
of the calibration and beam uncertainties is that the
combination $\ln {\cal C}_{10} + \sum_i
\ln A_i^2 - \langle(\ell+\frac{1}{2})^2\rangle_B \Delta (\sigma_{bi})^2$ is
adjusted for each grid parameter combination to give the best fit with
errors. Marginalization over the continuous parameters is done by
calculating a Fisher matrix, and assuming a Gaussian distribution in the
posterior distributions in the $\{ \ln {\cal C}_{10}, 
\ln A_i^2, \Delta (\sigma_{bi})^2\} $ variables. Marginalization over
the grid parameters is done by discrete integration. 

In addition to the parameters given in Table~\ref{paramlist},
we examine four ``derived'' parameters, which are functions of our basic 7
variables, $\{t_0$, $h$, $\sigma_8^2$, and $\Gamma_{\rm eff} \}$. 
Here $h$ is the value 
of the Hubble expansion parameter, $H_0 = 100 ~h$ km/s/Mpc
and is given by
\begin{equation}
h = {\sqrt \frac{\omega_b + \omega_{cdm}}{1 - \Omega_k - \Omega_\Lambda}} \ .
\end{equation}
The age of the universe is
\begin{equation}
t_0({\rm Gyr}) \approx \frac{9.778}{h} \int_0^1
\frac{2x^2}{\sqrt{\Omega_m+\Omega_kx^2+\Omega_{\Lambda}x^3}}dx \ ,
\end{equation}
where we have dropped the minor effects of relativistic particles
here, but not when we do our actual comparisons. 
The
variance in the (linear) density fluctuation spectrum on the scale of
clusters of galaxies ($8 h^{-1}$ Mpc) is $\sigma_8^2$.
The shape of the linear
density power spectrum is described by the parameter
\begin{equation}\begin{split}
&\Gamma_{\rm eff}=\Gamma + (n_s- 1)/2 ,\quad {\rm where}\\
&\Gamma = \Omega_{m}
\, h \, \, e^{-(\Omega_b(1+\Omega_{m}^{-1}\sqrt{2h}) -0.06)} \ .
\end{split}\end{equation}
Note that $\Gamma_{\rm eff} \approx \Omega_m h$ with 
corrections due to baryon density
and spectral tilt $n_s$ over the region probed by large scale
structure observations that approximate the dominant dependences. 
The derived parameters are not marginalized over, and do not define
dimensions in the parameter grid because they are determined given a set of
parameters $\vec{y}$.
The derived parameters are calculated at each
point on the model grid.  We can make Bayesian estimates of
those parameters 
using equation~\eqref{prior} with $X$ being the derived
parameter of interest.

We can also use the derived parameters to account for 
other cosmological information.  Estimates of
these parameters from other, non-CMB, observations can be 
included as ``prior constraints". Equation~\eqref{prior}
is cast in a
form that gives a likelihood as a function of our 
parameters, and priors $P_{prior}(\vec{y})$.

All our analyses include the loose prior implicitly
imposed by the edges of the database given in Table~\ref{paramlist},
and two other very weak priors that are generally accepted by most
cosmologists:  models are restricted to those where the 
current age of the universe is $t_0 > 10$Gyr, and 
where the matter density $\Omega_m$ is greater than 0.1.

We add to these constraints a series of 
additional priors:
\begin{itemize}

\item {\bf weak-h :} $0.45 < h < 0.90$.
This is a tophat restriction, designed to allow the CMB data,
rather than arguable priors, to drive the results.

\item {\bf LSS :}  We employ
two constraints based on large scale structure (LSS) observations, on
the combinations $\sigma_8 \Omega_m^{0.56}$ and $\Gamma_{\rm eff}$.
For both priors, we adopt a Gaussian distribution convolved with a
top-hat distribution, characterized by the parameters $\sigma_8
\Omega_m^{0.56} = 0.47_{-.02, -.08}^{+.02, +.11}$ and $\Gamma_{\rm
eff} = 0.21_{-.03, -.08}^{+.03, +.08}$, where the first error gives
the 1-$\sigma$ point of the Gaussian distribution, and the second
error gives the extent of the tophat distribution. Our basic
philosophy is to adopt priors that are not overly restrictive since
the LSS data is still improving. The motivation for the choice and the
discussion of the LSS data is given in \citet{bond02b}. The
$\Gamma_{\rm eff}$ distribution encompasses recent results from the
2dF and SDSS surveys.  The $\sigma_8$ distribution encompasses results
from recent weak lensing surveys. It also covers many of the cluster
abundance determinations using X-ray temperature and other cluster
data.

\item {\bf LSS($low$--$\sigma_8$):} There are currently a few cluster
abundance estimations that point to values of $\sigma_8$ that are
lower than the weak lensing estimates. Although our standard LSS prior
takes most of these variations into account by its spread, we have
also tested the effect of shifting the entire $\sigma_8$-distribution
downward by 15\%, to $\sigma_8 \Omega_m^{0.56} = 0.40_{-.02,
-.08}^{+.02, +.11}$. We keep the $\Gamma$ prior the
same.

\item {\bf HST-h :}  We strengthen our $h$ prior,
based on the Hubble Space Telescope Key Project
measurement~\citep{freedman00} of the Hubble constant, $h = 0.72 \pm 0.08$.  
This is a Gaussian prior with the stated error as the 1--$\sigma$ points.  

\item {\bf Strong data :}  We combine
the LSS prior given above with the HST-$h$ prior, and add
a constraint (in the $\Omega_{\Lambda}$ vs. $\Omega_{tot}$ plane)
based on surveys of the brightness vs. 
redshift relation of Type 1a Supernovae~\citep{perlmutter99,riess98}.

\item {\bf Flat :}  Inflation models generally predict a flat
geometry, and recent evidence supports 
this~\cite[eg][]{debernardis00,halverson01}.  For the converted,
we investigate the effects of holding $\Omega_{tot}$ equal to one. 
\end{itemize}

\section{Constraints on Cosmological Parameters from CMB Spectra}
\label{params}

We apply these methods to three combinations of CMB data, using a 
series of priors for each case.  By doing so, we can investigate the 
power of adding \acbarsp data to the current cosmological mix, as well 
as the dependence of results on the strength and nature of the 
applied priors.

The full COBE-DMR power spectrum of~\citet{bond98} is included in
all our analyses.
The first combination of CMB data, \acbar+DMR, investigates the 
potential for using 
the DMR low-$\ell$ anchor with the 
damping-tail measurement of \acbarsp as an independent check 
on previous CMB-based cosmological parameter estimation.

\begin{figure*}[ht!]
\resizebox{\hsize}{!}{
  \rotatebox{-90}{
    \includegraphics{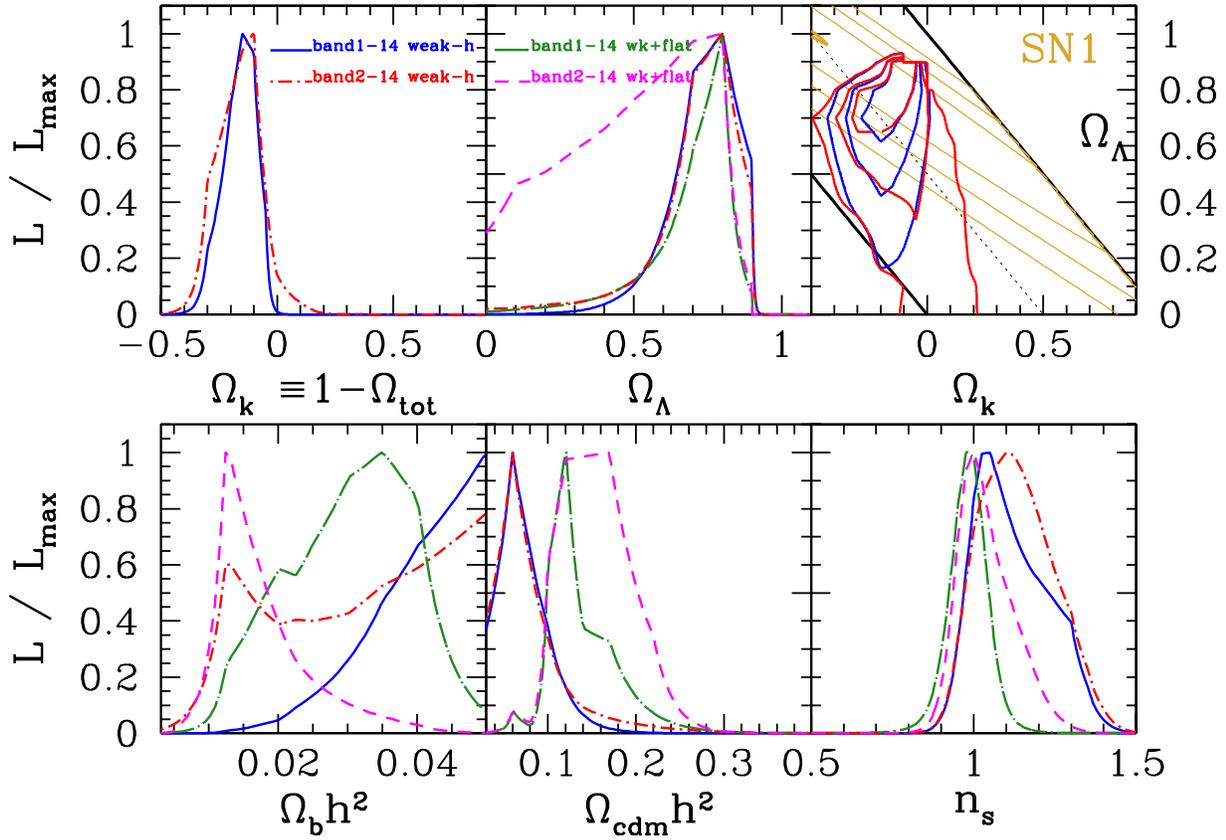}}}
\caption{\protect\small
Likelihood curves for \acbar+DMR with the $weak$ and $weak$+flat priors.
Each prior case is plotted with and without the first \acbarsp band
(centered on $\ell = 187$) included in the analysis.  In the upper--right
panel, the 1 to 3-$\sigma$ contours are shown for the 2D
$\Omega_k$--$\Omega_\Lambda$ likelihood with the $weak$ prior and
bands 1-14 (blue) and bands 2-14 (red).
The thick black lines define $\Omega_m=0$ and $\Omega_m=1$ and the dotted
black line defines $\Omega_m=0.5$.
The yellow contours are the 1, 2 and 3-$\sigma$ levels of constraints based
on Type 1a Supernovae. The lack of
stability of the curves (for $\omega_b$ in particular) indicates that
\acbar+DMR alone is not sufficient for robust parameter estimation.}
\label{fig:1DLacbar}
\end{figure*}

Figure~\ref{fig:1DLacbar} shows marginalized likelihood 
curves for \acbarsp and DMR only, using the weak-$h$ and weak-$h$+flat priors.  
These are displayed for
each of two sets of \acbarsp data, the first being the full
spectrum (bands 1-14), the second being bands 2-14.  
The curves show the parameter estimates are very prior-dependent, 
and also depend upon whether we include the lowest $\ell$ band data,
despite the 
large errors in that band. These variations are a result of \acbarsp
not pinning down the peak/dip structure well. Parameter estimates are
then relying more on the specific shape of the damping tail. 

The physics of damping is well known and clearly depends upon the
cosmological parameter characterizing the strength of the viscous and
diffusive couplings, $\omega_b$. It might then be thought that by using
only \acbarsp and DMR we could get a strong constraint on this
parameter.  However, when we take account
of all of the influences that determine the damping scale in
$\ell$-space, it is found to be relatively insensitive (see \eg
\citet{sievers02} for a discussion). Once some information is given
on the peak/dip structure that helps to pin the parameters, \acbarsp
improves the quality of the determinations by virtue of its small
error bars in the damping tail region. We therefore proceed 
by including CMB data in the $\ell$ range between DMR and \acbar,
over the first three acoustic peaks.

Many CMB experiments have made sensitive measurements of
the power spectrum in the region of those peaks;  we
choose here to combine the low-$\ell$ measurement of DMR with a 
set of recent higher-$\ell$ observations, comprised of 
Archeops~\citep{benoit02}, Boomerang~\citep{ruhl02},
DASI~\citep{halverson01}, MAXIMA~\citep{hanany00}, 
and VSA~\citep{scott02}.  We also include the recent
high-$\ell$ results of CBI~\citep{pearson02}.
We give the label ``Others" to this aggregate set of data 
and first investigate the parameter
extraction that can be done with these measurements,
{\em sans} \acbar.  Finally, we investigate the 
improvements made when adding \acbarsp to the mix, in a combination
we label ``\acbar+Others".

\begin{figure*}[ht!]
\resizebox{\hsize}{!}{
  \rotatebox{-90}{
    \includegraphics{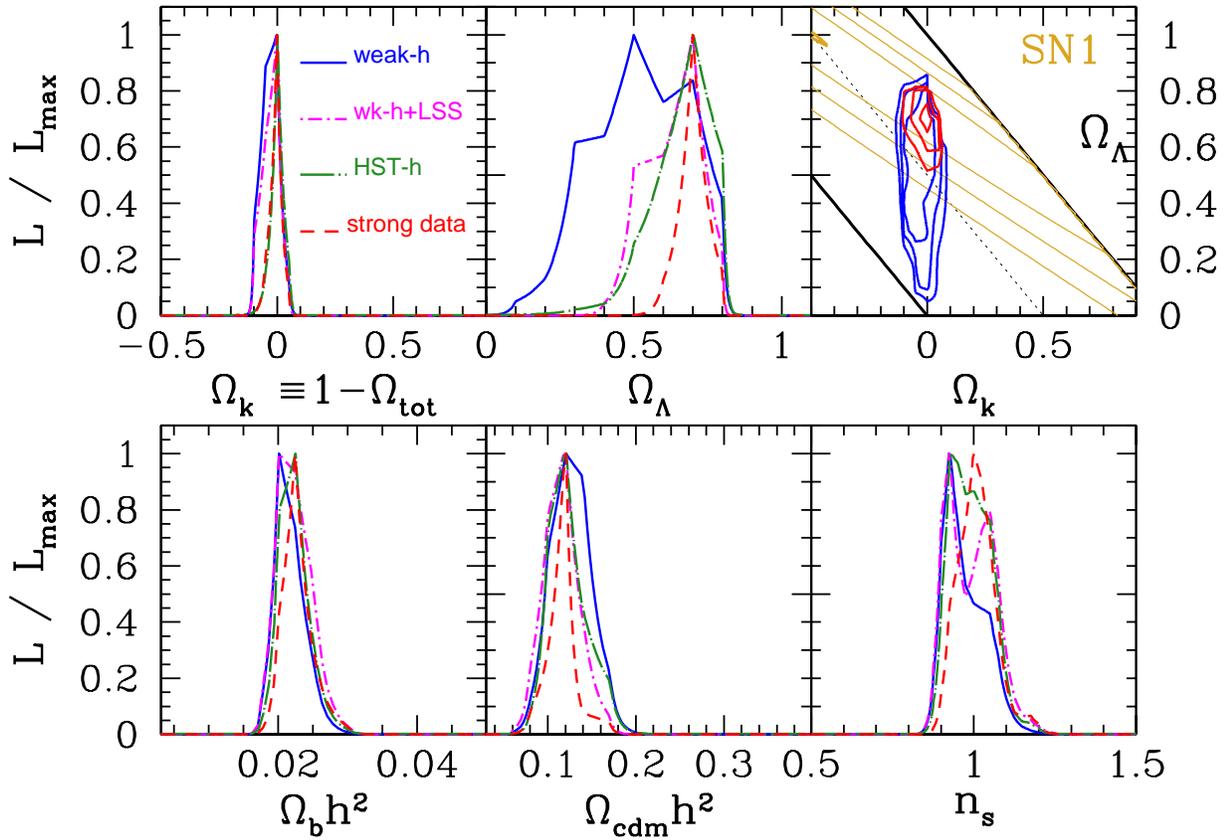}}}
\caption{\protect\small
Likelihood curves for Archeops+B98+CBI+DASI+DMR+MAXIMA+VSA
(``Others'') with the $weak$, $weak$+LSS, HST-$h$, and {\em strong data}
priors.  The $\Omega_k$--$\Omega_\Lambda$ contours are shown for the 
$weak$ (blue) and {\em strong data} (red) cases.
The yellow contours are the 1, 2 and 3-$\sigma$ levels of constraints based
on Type 1a Supernovae.
CMB estimates of $\Omega_k$, $\omega_{cdm}$, and $\omega_b$ are
stable with sensible behavior as additional 
priors are employed.  }
\label{fig:1DLothers}
\end{figure*}

\begin{figure*}[ht!]
\resizebox{\hsize}{!}{
  \rotatebox{-90}{
    \includegraphics{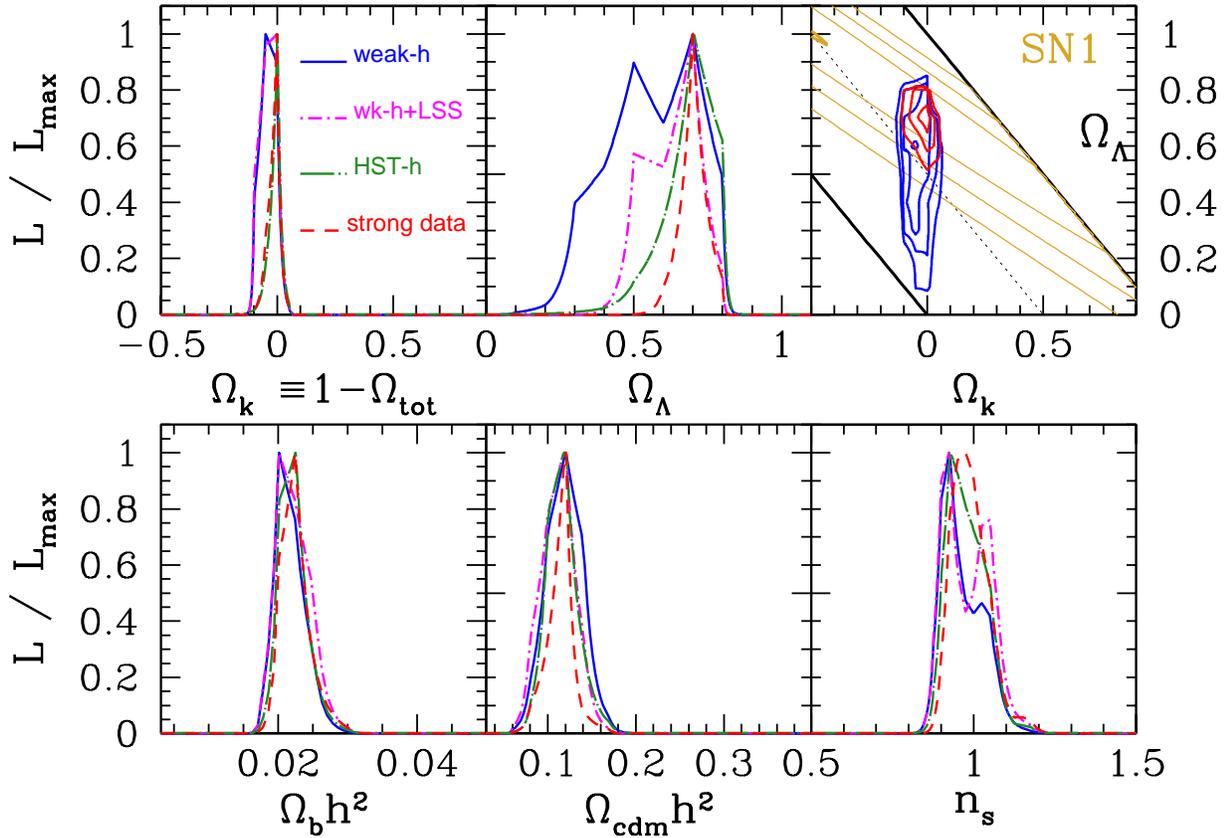}}}
\caption{\protect\small
Likelihood curves for \acbarsp+Archeops+B98+CBI+DASI+DMR+MAXIMA+VSA
(\acbar+``Others'') with the $weak$, $weak$+LSS, HST--$h$, and
{\em strong data} priors.    The $\Omega_k$--$\Omega_\Lambda$ contours
are shown for the 
$weak$ (blue) and {\em strong data} (red) cases. 
The yellow contours are the 1, 2 and 3-$\sigma$ levels of constraints based
on Type 1a Supernovae.
The positions and widths of these curves do 
not differ significantly
from those in Figure~\ref{fig:1DLothers} despite the addition
of the low noise \acbarsp data through the damping tail.  
A comparison of the curves here containing the LSS prior
($weak$+LSS and {\em strong data}) with those derived using the 
a lower estimate (discussed in the text) shows only small
changes.  The most noticeable changes are an upward shift in the 
lower tail on $\Omega_\Lambda$,
and a broader and higher--value $n_s$ peak.
Table~\ref{estimatetab} gives numerical estimates of
these parameters, derived by integration of these curves.}
\label{fig:1DLall}
\end{figure*}

Two plots of one-dimensional marginalized likelihood curves,
for ``Others" and ``\acbar+Others", are given in 
Figures~\ref{fig:1DLothers} and \ref{fig:1DLall} respectively.
Here we see that results are generally stable to the application
of priors - that is, the application of a prior may narrow the result,
but does not move it outside the range imposed by other
priors. 

It is remarkable how many parameters are well constrained 
in Figures~\ref{fig:1DLothers} and \ref{fig:1DLall}.  Using only
a weak-$h$ prior, four of the five parameters shown 
($\Omega_k$, $\Omega_b h^2$, $\Omega_{cdm} h^2$, and $n_s$) 
are well localized.  The detection of 
$\Omega_\Lambda$ becomes stronger as stronger priors are applied.  

We have investigated the effects of dropping the first
\acbarsp bin on the results shown in 
Figure~\ref{fig:1DLall}, and unlike the 
\acbar+DMR results,
the effects of this action are negligible.  This is not surprising, given 
that the first peak information in ``Others" and ``Others+\acbar"
is driven by the other measurements.  

The likelihood curves shown in these figures can be used to find 
confidence intervals on each parameter for each prior.  We can proceed
in a similar way to constrain the values of other parameters, such
as $h$ and the current age of the universe ($t_0$).  
Table~\ref{estimatetab} gives the median values and
1-$\sigma$ limits (16\% and 84\% integrals of the likelihood)
for a set of cosmological parameters.
For $\tau_C$, the 95\% confidence upper limit is given.  
A comparison of the values in the Table, or of 
Figures~\ref{fig:1DLothers} and \ref{fig:1DLall}, 
shows that the impact
of adding \acbarsp to this mix is not dramatic.  There are 
modest improvements in some parameter estimates, most notably 
$\Omega_\Lambda$ and $\Omega_{cdm} h^2$.

Not immediately obvious from the figures is the 
improved rejection of $\Omega_\Lambda = 0$ models.
One measure of this is the improvement in the 3-$\sigma$ lower
limit on $\Omega_\Lambda$, found by integrating the likelihood
curve.  This 3-$\sigma$ limit improves from $\Omega_\Lambda > 0.086$
for ``Others" to $\Omega_\Lambda > 0.136$ for ``\acbar+Others"; that
is, the probability of having a lower value of $\Omega_\Lambda$ than
these is $0.14\%$.  This depends upon the specific
range in $h$ we are allowing in our weak prior. 

The source of this rejection of models near $\Omega_\Lambda = 0$ 
can be illustrated by examining the $\chi^2$ of the aggregate dataset 
to the best-fit models, in both the $\Omega_\Lambda = 0$ and 
the ``free" $\Omega_\Lambda$ cases.
We find, for the ``\acbar+Others" dataset (consisting of 116 band powers)
$\chi^2 = 140$ and $\chi^2 = 160$ for the best-fit ``free $\Omega_\Lambda$"
and $\Omega_\Lambda = 0$ models, respectively.  Thus, while both models
plotted in Figure~\ref{fig:aps} appear reasonable to the eye, the
fit is significantly better for the $\Omega_\Lambda = 0.5$ model.

Unfortunately, calculating the effective number of degrees of freedom
in this $\chi^2$ is not straightforward. Taking beam and calibration 
uncertainties for each observation as a total of 16 parameters 
in the fit added to the 7 cosmological parameters, we know the 
effective degrees of freedom lies in the 
range $93 < \mbox{dof} < 116$.  Adopting 
dof=100 as a reasonable estimate,  the probability of 
finding $\chi^2 > 140$ and $\chi^2 > 160$ 
are $P_> = 0.0051$ and 0.00013 respectively.  
We caution the reader against strict interpretation of these 
statistically high $\chi^2$ in the face of this 
very heterogeneous data set.  Instead, we note the significant
improvement in $\chi^2$ enabled by the addition of a single
parameter in the fit.

Interestingly, the $\chi^2$ difference of the best-fit models
is roughly consistent with the ratio of the marginalized likelihoods
(for the weak-$h$ prior) at the peak of the likelihood curve
near $\Omega_\Lambda = 0.7$ and its level at $\Omega_\Lambda = 0$.
Exact correspondence would be expected if our parameter
likelihood function had a Gaussian form.

One would think that the information \acbarsp adds at high $\ell$
would significantly improve the determinations of $\omega_b$ because
of viscous damping and $n_s$ just because of the increased
$\ell$-baseline. This is clearly not the case.  We have discussed the
damping tail issue already. The near degeneracies due to correlations
among certain parameter combinations imply that increased data does
not necessarily lead to increased precision on the cosmological
parameters.  It is likely that the lack of improvement in these
projections to individual parameters is due to degeneracies in the
parameter space.

Such degeneracies are well known and have been discussed at length
in the literature;  see \citet{efstathiou99} for an extensive treatment
of some of the most pernicious of these.  For example, one
of these (the ``geometric degeneracy"), leads to nearly identical
angular power spectra for particular combinations of 
($\Omega_\Lambda$, $\Omega_k$, $\Omega_m$, $h$), while leaving $\omega_b$
and $\omega_{cdm}$ fixed.  The breaking of this geometric degeneracy 
is the reason estimates of $\Omega_\Lambda$ and, to a lesser
extent $\Omega_{cdm} h^2$, improve so dramatically as 
stronger priors on $h$ are applied.  Other less exact degeneracies,
such as one between amplitude, $n_s$ and $\tau_C$, lead to similar 
broadenings of these projected likelihood curves.

By choosing our database parameters well (\eg using the physical
densities $\omega_b$ and $\omega_{cdm}$ rather than the densities
relative to critical) we have minimized the effects of some 
potential degeneracies.  By exploring the parameter eigenmodes 
we can escape the limitations
of the canonical parameters and determine the true power of any
combination of datasets.  This process is in fact quite familiar
to most cosmologists;  the Type 1a supernovae results are often
considered as limiting the parameter ``eigenmode"
$\Omega_\Lambda$--$\Omega_{M}$.

Table~\ref{tab:eigenmodes} lists the best-determined five (of 
seven total) eigenmodes
for the ``Others" and ``\acbar+Others" analyses.  The coefficients 
describing those modes, and the errors on the eigenvalues,
are determined by ensemble averages of the likelihood derivatives 
over the parameter database.  We introduce the probability-weighted 
ensemble average of a parameter $y_a$,
\begin{equation}
  \left< y_a \right> = 
  	\int y_a P_{prior}(\vec{y}) {\cal L}(\vec{y}) d\vec{y} \ ,
\end{equation}
and the probability weighted ensemble average of the differentials
\begin{equation}
  \left< \delta y_a \, \delta y_b\right> = 
  \int \delta y_a \delta y_b P_{prior}(\vec{y}) {\cal L}(\vec{y}) d\vec{y} \ ,
\end{equation}
where $\delta y_a \equiv y_a - \left< y_a \right>$.

The eigenmodes themselves ($\xi_k$) and their 
errors ($\sigma^\xi_k$) are given by the application of
a rotation matrix ${\bf R}$ to the parameter vector $\vec{y}$,
\begin{eqnarray}
\xi_k &=& \sum_a R_{ak} \delta y_a \ ,  \\
\left<\delta y_a \delta y_b\right> &=&
  \sum_k R_{ak} (\sigma^\xi_k)^2 R_{bk} \ .
\end{eqnarray}
In this eigenmode analysis, we have used the fractional deviations
$\delta \omega_b /\left<\omega_b\right>$ and 
$\delta \omega_{cdm} /\left<\omega_{cdm}\right>$ 
as parameters, rather than $\omega_b$ and $\omega_{cdm}$, to set their
deviation magnitudes on more equal footing with those of the other
parameters. 

Inspection of the table shows that while the eigenmodes 
for ``Others" and ``\acbar+Others" are not identical, they are
very similar.  In most cases they are dominated by contributions from
one or two cosmological parameters, but in all cases there 
are significant components from several parameters.
The table shows much more clearly the impact of adding 
\acbarsp to the dataset; all of the eigenvalue uncertainties improve, 
but the greatest improvement is in the fifth eigenmode, which becomes
the fourth one when \acbarsp is added. It is dominated by the 
cosmological constant. 

\section{Constraints on $\sigma_8$ from the Sunyaev-Zeldovich Effect}

The \acbarsp results provide the first data above 
$\ell=2000$ at 150~GHz.
The recent CBI Deep field results~\citep{mason02}, at 30~GHz,
have indicated a possible excess over the expected 
primary anisotropy signal
at $\ell > 2000$. The most promising
candidate for the source of the excess is the Sunyaev-Zeldovich Effect
(SZE) due to the scattering of CMB off hot electrons in the
intra-cluster medium (see \citet{birkinshaw99} for a recent
review). The CBI results have been interpreted in the context of the
SZE with tentative constraints being obtained on the value of
$\sigma_8$ \citep{bond02b,komatsu02wc,holder02}. 
The BIMA array \citep{dawson02}, operating
at 30~GHz, has also reported detection of power at higher $\ell$
which also has been attributed to the SZE by \citet{komatsu02wc}. 

Parameter fitting using secondary effects such as the
SZE must be approached with caution.  Both numerical and analytical
predictions for the SZE power spectrum suffer from a number of
uncertainties.  The results of different simulations,
although in general agreement, show significant differences in both
the amplitude and shape of the predicted spectrum.  Analytical models
suffer from uncertainties inherent in modeling the profile of the
clusters.  In addition, cooling and heating effects
in the clusters are not yet well understood and most simulations and
analytical models do not take these effects into account. 
Simulations have shown that for modest deviations about the
concordance $\Lambda CDM$ model, the SZE
angular power spectrum scales as ${\cal C}^{\rm SZ}_{\ell} \sim (\Omega_b
h)^2 \sigma_8^7$. 
Despite uncertainties in the physics of cluster models, especially the
role of energy injection, and the relatively large errors on the 
observations, this very 
strong dependence of the SZE spectrum on $\sigma_8$
enables the derivation of significant constraints on 
$\sigma_8$ from current data.

We choose to model the
SZE using two angular power spectrum templates.  
The first is obtained from 
large Smoothed Particle Hydrodynamics (SPH)
simulations of $\Lambda CDM$~\citep{bond02a,wadsley02}.
The second is obtained from an
analytical model (see \citet{zhang02, bond02b} for details).  Both
templates were scaled to a fiducial value of $\Omega_b h =
0.035$. 

Although the power in the primary spectrum is falling rapidly compared
to the rising contribution of the SZE at $\ell > 2000$, interpretation of
the low noise \acbarsp band powers around the cross-over region 
is sensitive to
the contribution of the primary signal together with
the secondary. 
Rather than consider a full range of parameter space
we select a simple model for the primary
spectrum, a best fit flat model for the 
``\acbar+Others" data combination ($\Omega_b=0.047$, $\Omega_{cdm} = 0.253$,
$\Omega_{\Lambda}=0.7$, $h=0.69$, $n_s=0.975$ and $\tau_C=0.2$).  The
primary model was normalized with the best fit amplitude obtained from
the fits. However, uncertainties in the model parameters
affect the overall amplitude of the primary spectrum at scales
$\ell\approx 2000$ where the primary and secondary signals are
comparable. We chose to parameterize the freedom in the primary and
secondary spectra by two {\it effective}
parameters, an amplitude in the primary power spectrum $q_{2K}^{\rm eff}$
and a scaling factor for the SZE $\sigma_8^{\rm SZ}$. Uncertainty in
the primary parameter $q_{2K}^{\rm eff}$ represents the
uncertainties in
a number of dominant effects given by the combination of
parameters: $\sigma_8$, $\tau_C$, and $n_s$, as shown in
section~\ref{params}. Most significantly,
it also incorporates the effect of systematic uncertainties such as an
overall calibration and beam uncertainty in the data. The secondary
amplitude parameter $\sigma_8^{\rm SZ}$ describes the scaling of the SZE
spectrum and can be related to $\sigma_8$ via $\sigma_8^{\rm SZ}
\approx (\Omega_b h/0.035)^{0.29} \sigma_8$.

We select points with $\ell>1500$ for fitting and use the offset
lognormal approximation as described in Section~\ref{pssec}. The target
model is now $Z^T_B = \ln({\cal C}_B+g_{\nu}{\cal C}_B^{\rm SZ} + x_B)$ where
the primary and SZE spectra have been filtered by the appropriate
window functions for each band power and $g_{\nu}$ is the frequency dependent
scaling of the SZE (a factor of $\sim 4$ in
power lower at 150GHz compared to 30GHz). Top--hat windows are used to
model the BIMA band powers. The primary model is scaled by
$q_{2K}^{\rm eff}$ over the range $(0.1,1.8)$ 
and the secondary model
is scaled by $\sigma_8^{\rm SZ}$ over the range $(0.5,1.4)$.
At each point in the grid, we calculate the quantity $-2\ln(L)$
for the model.

To assign a realistic uncertainty to the amplitude of
the primary models, we add a Gaussian prior in the amplitude
$q_{2K}^{\rm eff}$ with a width of 20\% RMS. This is chosen to
reflect the uncertainty in the amplitude of
the best fit models obtained in the parameter fits described in
section~\ref{params}. As an example, the \acbarsp data fixes the
amplitude of our template model with an RMS of 17\% while the CBI Deep
field data fixes the same model with an RMS of 21\%. We check that
our results are robust to a change in the width of the prior by
fitting with a 10\% and 40\% RMS width.  The marginalized, best fit
value, and upper errors for $\sigma_8^{\rm SZ}$ reported below changes
by only 0.1\% while the lower error changes by 1\% on average.

\begin{figure*}[ht!]
\resizebox{\hsize}{!}{
  \includegraphics{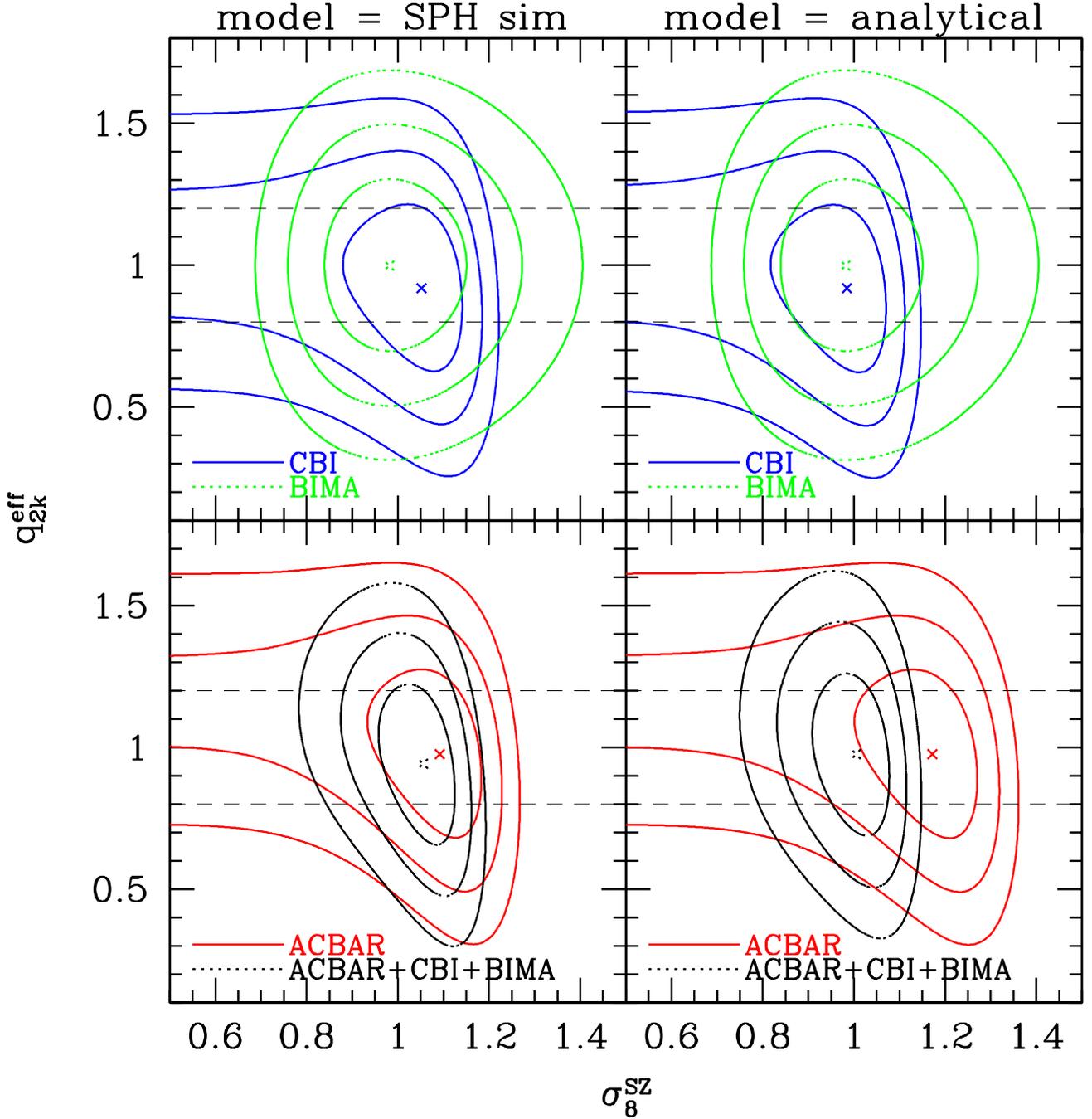}}
\caption{\protect\small 1, 2, and 3--$\sigma$ contours for various
combinations of data sets in the
$(q_{2K}^{\rm eff},\sigma_8^{\rm SZ})$ plane. The left panels show
fits obtained using the SPH template and the right panels are the equivalent
for the analytical model. Both CBI and \acbarsp data points constrain the
upper values of $\sigma_8^{\rm SZ}$ but not the lower values as they are
sensitive to the amplitude of the primary spectrum. The higher--$\ell$ BIMA
observations are insensitive to the primary component and therefore provide
a strong lower bound. The combination of the three datasets (bottom
row -- dotted, black contours) show a strong detection of the SZ component.
The dashed parallel lines show the width of the Gaussian prior imposed on
$q_{2K}^{\rm eff}$. We use a lognormal distribution for the BIMA
band powers.}
\label{sz1}  
\end{figure*}

In Fig.~\ref{sz1} we show contour plots of the $\chi^2$ grids in the
($q_{2K}^{\rm eff}$, $\sigma_8^{\rm SZ}$) plane. We subtract the $\chi^2$
value at the minimum from the grids and the (2.3,
6.17, 11.8) contours give an indication of where the 1, 2, 3--$\sigma$
levels would fall if the likelihoods were Gaussian.  We show results for both
template models used in the analysis.  The contours show that the data
is only weakly dependent on the $q_{2K}^{\rm eff}$ amplitude
with a slight movement to higher $\sigma_8^{\rm SZ}$ for low
$q_{2K}^{\rm eff}$ as expected for \acbarsp and CBI data. This reflects
the fact that the only bands sensitive to the primary amplitude are
the lowest $\ell$ bands included from the \acbarsp and CBI data. The
high--$\ell$ BIMA data is fully degenerate in the amplitude of the
primary, as expected, and the contours in the $q_{2K}^{\rm
eff}$ direction simply reflect the Gaussian prior. It does however
give a strong lower bound although this depends
strongly on the lognormality of the BIMA band power distribution. The $x_B$
offsets are not available for BIMA and are set to $x_B=0$ although we also
show results where no lognormal transformation was applied to the BIMA
points (Table~\ref{tabsz}). \acbarsp and CBI provide strong upper bounds.
A combination of the three data
sets give strong constraints in $\sigma_8^{\rm SZ}$ with the contours
showing a slight tilt in the expected direction with respect to
$q_{2K}^{\rm eff}$.

One major obstacle in using a secondary effect such as the SZE to fit
parameters is the non-Gaussian nature of the signal from non-linear
structures such as clusters. In general the non-Gaussianity will
increase the sample variance of the underlying signal. Treating the
data as Gaussian can therefore result in 
an overestimate of the significance of the constraint. 
This effect has been investigated using
numerical simulations of the SZE \citep{white02,zhang02} and also by
calculating the contribution to the covariance by the fourth order,
trispectrum term $T_{\ell\ell'}$ \citep{cooray01,komatsu02wc}. In general
the sample variance is found to be a factor $\sim 3$ higher than the
Gaussian equivalent with some dependence on $\ell$ and also on the
width of the bands being considered ($\ell$-$\ell'$ correlations are
also altered by the non-Gaussianity). In order to include this effect
in the errors and correlations we scale the inverse Fisher matrix of
the band powers as
\begin{equation}
{\cal \bar{F}}_{BB'}^{-1} \equiv f^{\rm ng}_{B}{\cal
F}_{BB'}^{-1}f^{\rm ng}_{B'} \ ,
\end{equation}
where $f^{\rm ng}_{B}$ is the scaling factor. Here we chose
$f^{\rm ng}_{B}$ such that the sample variance component of the error
is a factor of 3 larger than the expected Gaussian case which can be
approximately evaluated as
\begin{equation}
\Delta C_{\ell} \approx \frac{\sqrt{2}C_{\ell}}{\sqrt{f_{\rm sky}\Delta_B
(2\ell_b+1)}} \ .
\end{equation}

\begin{figure*}[ht!]
\resizebox{\hsize}{!}{
\includegraphics{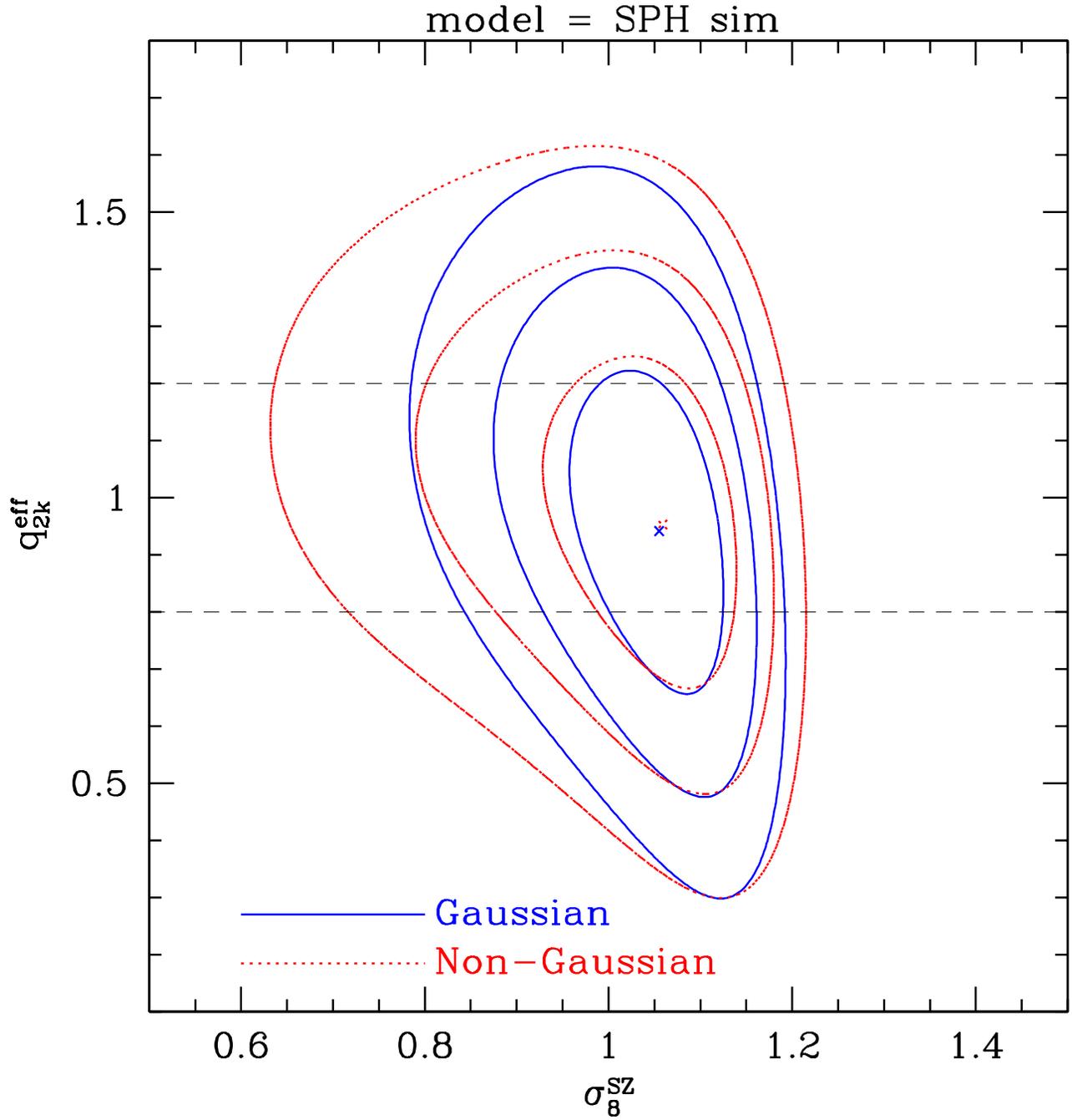}}
\caption{\protect\small The plot shows the effect of adding a correction
for the increased sample variance of the data due to the non--Gaussian
scatter. Both contours are for the combination \acbar+CBI+BIMA. The effect
is significant, particularly at the 3--$\sigma$ level.}
\label{sz2}  
\end{figure*}

Fig.~\ref{sz2} shows the effect of the correction on 
($q_{2K}^{\rm eff}$,$\sigma_8^{\rm SZ}$). In general the correction
changes from experiment to experiment and from band to band due to the
different sample variance component in the uncertainty of each band. It
is therefore important to include this effect in the analysis since it
can have a substantial effect on the structure of the contours as
opposed to a simple rescaling of the confidence limits. We find that
the correction has significant effects on the allowed region,
particularly at the 2 to 3-$\sigma$ level.

\begin{figure*}[ht!]
\resizebox{\hsize}{!}{
\includegraphics{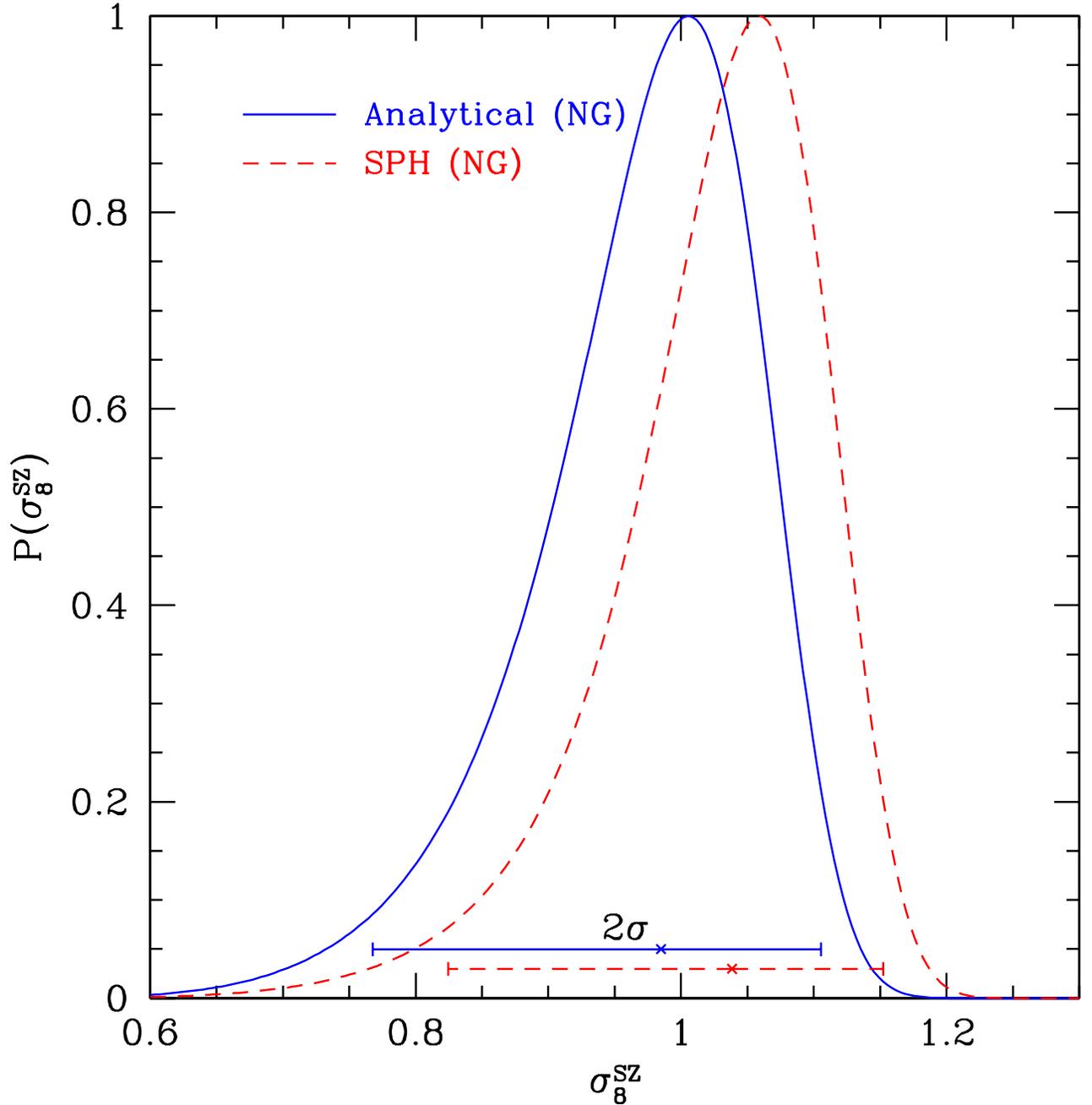}}
\caption{\protect\small Marginalized likelihoods for the \acbar+CBI+BIMA
combination. The results of fits using both the SPH (red, dashed) and
analytical (blue, solid) models are shown. Both cases include non--Gaussian
corrections. The 2--$\sigma$ upper and lower bounds and median values were
obtained by computing the 2.5\%, 97.5\%, and 50\% integrals of the
likelihoods respectively. We find that the SPH model prefers slightly higher
values for $\sigma_8^{\rm SZ}$. This is due to the fact that the SPH spectra
show less power than the analytical models on large scales.}
\label{sz3}  
\end{figure*}

\begin{figure*}[ht!]
\resizebox{\hsize}{!}{
\includegraphics{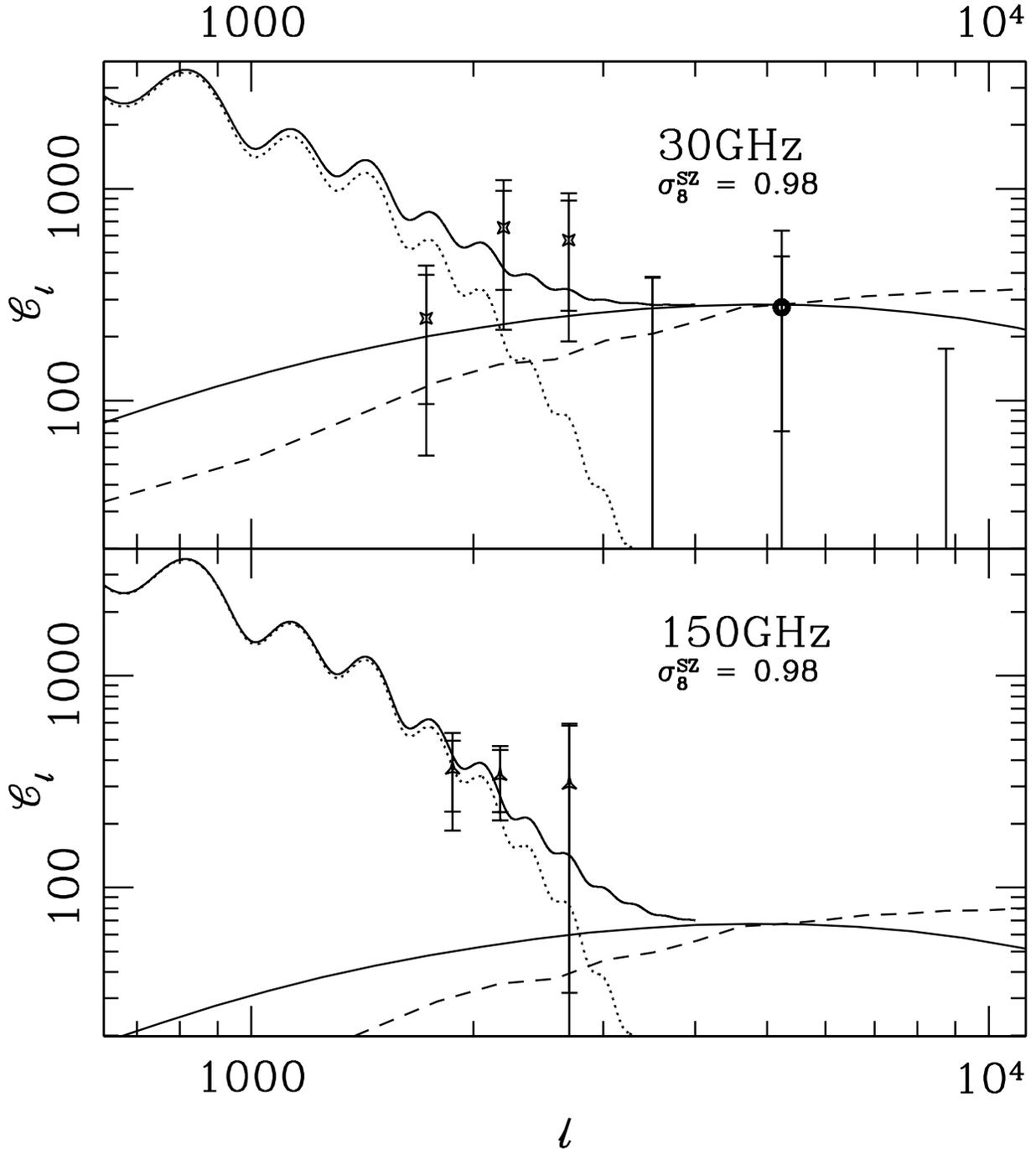}}
\caption{\protect\small Best fit primary+SZ model amplitudes. The top panel
shows the SZ model scaled to an observing frequency of 30GHz which
corresponds to the BIMA and CBI observations (first four rightmost points
and two leftmost points respectively). The lower panel is for 150GHz,
corresponding to the \acbarsp results. The solid lines show the analytical
model at each frequency and the total primary+SZ. The dotted line shows the
fiducial spectrum used to model the primary contribution at $\ell > 1000$.
The dashed line shows the SPH spectrum used as a template shape. The SZ
contributions are scaled to a value of $\sigma_8^{\rm SZ} = 0.98$ to show
the amplitude at the best fit value obtained from the \acbar+CBI+BIMA
combination using the analytic model as template (with non-Gaussian
corrections). The extended errorbars for each point show the corrections
due to the increase in sample variance expected from the non-Gaussian
 signal.}
\label{sz4}  
\end{figure*}

To obtain best estimates on the value of $\sigma_8^{\rm SZ}$, we
marginalize over the $q_{2K}^{\rm eff}$ direction to recover
the one dimensional likelihood in $\sigma_8^{\rm SZ}$. We show the
resulting likelihoods in Fig.~\ref{sz3} for both analytical and SPH
templates. Both results include the non-Gaussian correction discussed
above. The 2-$\sigma$ region and median values shown as errorbars
are obtained by calculating the $2.5\%$, $97.5\%$ and $50\%$
integrals of the likelihoods respectively. The results for both
templates with 2-$\sigma$ error estimates are summarized in 
Table~\ref{tabsz}. We find that fitting with the SPH template results
in values for $\sigma_8^{\rm SZ}$ about 6\% higher than when the
analytical model is used. This effect is due to the SPH model having a
lower amplitude at larger scales than the analytical model
(Fig.~\ref{sz4}).

In Fig.~\ref{sz4} we show the primary and SZE templates used, scaled
to the best fit value of $\sigma_8^{\rm SZ}=0.98$ (at 30GHz and
150GHz) for the analytical case shown in Fig.~\ref{sz3}. The total
primary+SZE (analytical) model is also shown together with the SPH
template scaled to the same parameters. We see that the method can
obtain a good fit between the data and primary+SZE model at both
observing frequencies. Fig.~\ref{sz4} also shows the non-Gaussian
corrections to each band power error. 

It is important to note that if there is a non-negligible SZE
component to the observed power, it may affect the parameter fits which
assume only a primary contribution. However, the relative contribution
to the \acbarsp band powers is very small, only approximately 15\% in
the last three bands. We do not expect this to have any significant
impact on the parameter estimates derived in this work.  As future
observations increase the accuracy in this region of the 
spectrum, a fully consistent approach to parameter fitting
will have to be adopted.  Such an approach would
simultaneously account for primary and secondary anisotropy,
fitting for all parameters.

\section{Conclusions}

The \acbarsp data, the most sensitive to date in the damping-tail
region, are in good agreement with predictions of flat $\Lambda$-CDM 
models with adiabatic initial perturbations.
Considering the \acbarsp data together 
with other recent CMB results, 
we find that the addition of a single parameter to the model,
$\Omega_\Lambda$, dramatically improves the best fit, with an 
improvement of $\Delta \chi^2 = 20$ upon adding that one parameter.
Using very weak cosmological priors 
($\Omega_m > 0.1$, age $> 10$Gyr, $0.45 < h < 0.90$), the 3-$\sigma$ 
lower limit on the cosmological constant rises to $\Omega_\Lambda > 0.136$ 
upon including the \acbarsp data.

We find that the addition of \acbarsp data to the current CMB set
does not lead to substantial improvements in the 1-$\sigma$ estimates
of the canonical cosmological parameters.  However, in an 
eigenmode analysis, the addition of \acbarsp data does 
improve the rotated parameter
uncertainties, indicating that in this case the lack of 
improved errors on the pure cosmological parameters is 
probably dominated by degeneracies between those parameters.  

We fit a SZE component to the data using \acbarsp and other
measurements at high-$\ell$.  Our estimates for the value of the
effective quantity $\sigma_8^{\rm SZ}=(\Omega_b h/0.035)^{0.29}\sigma_8$
show an improvement
over previous estimates using only CBI and BIMA observations. Although
CBI and \acbarsp alone do not provide lower bounds on $\sigma_8^{\rm
SZ}$, the combination of the two observations results in a detection
greater than 3-$\sigma$ which is independent of the BIMA lower bound.

Our fits show that at a fiducial value of $\Omega_b h=0.035$ the
central values for $\sigma_8$ are consistently higher than other
estimates obtained using cluster data, weak lensing surveys and
primary CMB observations (see
\citet{bond02b} for a recent survey of $\sigma_8$ estimates).  
However, the results overlap with most other
estimates at the 2-$\sigma$ level.
As an indication, our LSS prior, a smoothed tophat on the
combination $\sigma_8\Omega_m^{0.56}$, translates roughly into a
smoothed tophat of $\sigma_8=0.92^{+0.21}_{-0.15}$ at $\Omega_m=0.3$.
Any {\it statistical} inconsistency therefore, appears to be
mild.  Furthermore, {\it systematic} uncertainties in the estimates
have not yet been taken into account.  The difference displayed by 
the numerical
and analytical based results of about 10\% is indicative of the
agreement between the two methods for predicting the SZE power
spectrum \citep{komatsu02wc}. In addition, entropy injection may have a
significant effect on the SZE power spectrum. These effects would
change the shape of
the SZE template and impact directly on our determination of $\sigma_8$.  
Nevertheless, we conclude that the in the context of the phenomenological
models adopted in this work, the data is consistent with a SZE component
at $\sigma_8$ values near the high end of independent estimates.

\acknowledgments

The support of Center for Astrophysics Research in Antarctica (CARA) 
polar operations has been essential in the installation and operation of 
\acbarsp at the South Pole.
We thank Ue-Li Pen and Pengjie Zhang for the
use of their analytical SZE models, and
Percy Gomez and Kathy Romer for assistance with \acbarsp 
observation and timestream monitoring.
This work has been primarily supported by NSF Office of Polar
Programs grants OPP-8920223 and OPP-0091840.
Matt Newcomb, Jeff Peterson, and Chris Cantalupo acknowledge partial
financial support from NASA LTSA grant NAG5-7926.
Chao-Lin Kuo acknowledges support from a Dr. and Mrs. CY Soong fellowship
and Marcus Runyan acknowledges support from a NASA Graduate Student 
Researchers Program fellowship.
Research in Canada is supported by NSERC and the
Canadian Institute for Advanced Research. The computational facilities at
Toronto are funded by the Canadian Fund for Innovation.

\bibliographystyle{apj}
\bibliography{cmbr_jr}

\pagebreak

\begin{deluxetable}{lllllllllllll}
\tabletypesize{\scriptsize}
\tablecaption{Parameter Values Used in Grid}
\tablewidth{0pt}
\tablehead{\colhead{Parameter}
& \colhead{Values}
}
\startdata
$\Omega_k$
& $-0.5$
& $-0.3$
& $-0.2$
& $-0.15$
& $-0.1$
& $-0.05$
& $0$
& $0.05$
& $0.1$
& $0.15$
& $0.2$
& $0.3$
\\
$$
& $0.5$
& $0.7$
& $0.9$
\\
\\
$\Omega_{\Lambda}$
& $0$
& $0.1$
& $0.2$
& $0.3$
& $0.4$
& $0.5$
& $0.6$
& $0.7$
& $0.8$
& $0.9$
& $1.0$
& $1.1$
\\
\\
$\omega_{cdm}$
& $0.03$
& $0.06$
& $0.08$
& $0.10$
& $0.12$
& $0.14$
& $0.17$
& $0.22$
& $0.27$
& $0.33$
& $0.40$
& $0.55$
\\
$$
& $0.8$
\\
\\
$\omega_b$
& $0.003125$
& $0.00625$
& $0.0125$
& $0.0175$
& $0.02$
& $0.0225$
& $0.025$
& $0.03$
& $0.035$
& $0.04$
& $0.05$
& $0.075$
\\
$$
& $0.10$
& $0.15$
& $0.2$
\\
\\
$n_s$
& $0.5$
& $0.55$
& $0.6$
& $0.65$
& $0.7$
& $0.725$
& $0.75$
& $0.775$
& $0.8$
& $0.825$
& $0.85$
& $0.875$
\\
$$
& $0.9$
& $0.925$
& $0.95$
& $0.975$
& $1.0$
& $1.025$
& $1.05$
& $1.075$
& $1.1$
& $1.125$
& $1.15$
& $1.175$
\\
$$
& $1.2$
& $1.25$
& $1.3$
& $1.35$
& $1.4$
& $1.45$
& $1.5$
\\
\\
$\tau_C$
& $0$
& $0.025$
& $0.05$
& $0.075$
& $0.1$
& $0.15$
& $0.2$
& $0.3$
& $0.4$
& $0.5$
& $0.7$
\\
\enddata
\tablecomments{\small
Grid point values for the six cosmological parameters that are 
varied discretely.
The grid is not spaced evenly for {$\omega_{cdm}, \omega_b, n_s, {\rm and}\;
\tau_C$}; points are purposefully more concentrated in 
regions in which the likelihood (found from previous datasets) is high.
We only calculate models on this grid which have $\Omega_m > 0.1$;
this, along with the edges of each parameter range, forms an implicit
prior in our analysis.
}
\label{paramlist}
\end{deluxetable}

\begin{deluxetable}{llllllllllll}
\tablecolumns{12}
\tabletypesize{\scriptsize}
\tablecaption{Parameter Estimates and Errors}
\tablewidth{0pt}
\tablehead{
\colhead{Priors}
& \colhead{Run}
& \colhead{$\Omega_{tot}$}
& \colhead{$n_s$}
& \colhead{$\Omega_bh^2$}
& \colhead{$\Omega_{cdm}h^2$}
& \colhead{$\Omega_{\Lambda}$}
& \colhead{$\Omega_m$}
& \colhead{$\Omega_b$}
& \colhead{$h$}
& \colhead{Age}
& \colhead{$\tau_C$}
}
\startdata
\multicolumn{2}{l}{weak--$h$} \\
& Others
& $1.03^{0.05}_{0.04}$ 
& $0.96^{0.09}_{0.05}$ 
& $0.022^{0.003}_{0.002}$ 
& $0.13^{0.03}_{0.03}$ 
& $0.53^{0.18}_{0.19}$ 
& $0.50^{0.19}_{0.19}$ 
& $0.072^{0.023}_{0.023}$ 
& $0.57^{0.11}_{0.11}$ 
& $14.9^{1.3}_{1.3}$ 
& $<0.48$
\vspace{2pt}
\\
& \acbar+Others
& $1.04^{0.04}_{0.04}$ 
& $0.95^{0.09}_{0.05}$ 
& $0.022^{0.003}_{0.002}$ 
& $0.12^{0.03}_{0.03}$ 
& $0.57^{0.16}_{0.18}$ 
& $0.47^{0.18}_{0.18}$ 
& $0.071^{0.022}_{0.022}$ 
& $0.57^{0.11}_{0.11}$ 
& $15.1^{1.3}_{1.3}$ 
& $<0.47$ 
\\
\multicolumn{2}{l}{HST--$h$} \\
& Others
& $1.00^{0.03}_{0.03}$ 
& $0.99^{0.07}_{0.07}$ 
& $0.022^{0.003}_{0.003}$ 
& $0.12^{0.03}_{0.02}$ 
& $0.68^{0.09}_{0.12}$ 
& $0.33^{0.11}_{0.11}$ 
& $0.049^{0.013}_{0.013}$ 
& $0.68^{0.08}_{0.08}$ 
& $13.7^{1.0}_{1.0}$ 
& $<0.45$ 
\vspace{2pt}
\\
& \acbar+Others
& $1.00^{0.03}_{0.02}$ 
& $0.97^{0.07}_{0.06}$ 
& $0.022^{0.003}_{0.002}$ 
& $0.12^{0.02}_{0.02}$ 
& $0.70^{0.07}_{0.10}$ 
& $0.31^{0.10}_{0.10}$ 
& $0.049^{0.013}_{0.013}$ 
& $0.68^{0.08}_{0.08}$ 
& $13.9^{0.9}_{0.9}$ 
& $<0.43$ 
\\
\multicolumn{2}{l}{wk--$h$+flat} \\
& Others
& (1.00) 
& $0.95^{0.08}_{0.05}$ 
& $0.022^{0.003}_{0.002}$ 
& $0.13^{0.03}_{0.03}$ 
& $0.59^{0.15}_{0.23}$ 
& $0.43^{0.19}_{0.19}$ 
& $0.056^{0.014}_{0.014}$ 
& $0.63^{0.10}_{0.10}$ 
& $13.9^{0.5}_{0.5}$ 
& $<0.34$ 
\vspace{2pt}
\\
& \acbar+Others
& (1.00) 
& $0.95^{0.07}_{0.05}$ 
& $0.022^{0.002}_{0.002}$ 
& $0.13^{0.02}_{0.02}$ 
& $0.66^{0.10}_{0.16}$ 
& $0.35^{0.15}_{0.15}$ 
& $0.049^{0.011}_{0.011}$ 
& $0.67^{0.09}_{0.09}$ 
& $13.8^{0.4}_{0.4}$ 
& $<0.31$ 
\\
\\[0.01cm]\tableline\\[0.01cm]
$$
& \acbar+Others
\\
\multicolumn{2}{l}{wk--$h$+LSS}
$$
& $1.03^{0.05}_{0.04}$ 
& $0.98^{0.09}_{0.07}$ 
& $0.022^{0.003}_{0.003}$ 
& $0.11^{0.02}_{0.03}$ 
& $0.64^{0.08}_{0.12}$ 
& $0.41^{0.11}_{0.11}$ 
& $0.067^{0.019}_{0.019}$ 
& $0.59^{0.09}_{0.09}$ 
& $15.2^{1.4}_{1.4}$ 
& $<0.51$ 
\vspace{2pt}
\\
\multicolumn{2}{l}{wk--$h$+flat+LSS}
$$
& (1.00) 
& $0.94^{0.07}_{0.05}$ 
& $0.022^{0.002}_{0.002}$ 
& $0.13^{0.02}_{0.02}$ 
& $0.65^{0.08}_{0.11}$ 
& $0.36^{0.10}_{0.10}$ 
& $0.050^{0.008}_{0.008}$ 
& $0.66^{0.07}_{0.07}$ 
& $13.9^{0.4}_{0.4}$ 
& $<0.32$ 
\vspace{2pt}
\\
\multicolumn{2}{l}{wk--$h$+flat+LSS($low$--$\sigma_8$)}
$$
& (1.00) 
& $0.98^{0.07}_{0.06}$ 
& $0.022^{0.002}_{0.002}$ 
& $0.12^{0.02}_{0.02}$ 
& $0.71^{0.06}_{0.07}$ 
& $0.28^{0.07}_{0.07}$ 
& $0.044^{0.006}_{0.006}$ 
& $0.71^{0.07}_{0.07}$ 
& $13.7^{0.4}_{0.4}$ 
& $<0.34$ 
\vspace{2pt}
\\
\multicolumn{2}{l}{{\em strong data}}
$$
& $1.01^{0.03}_{0.02}$ 
& $0.99^{0.07}_{0.05}$ 
& $0.023^{0.003}_{0.002}$ 
& $0.12^{0.02}_{0.02}$ 
& $0.70^{0.05}_{0.05}$ 
& $0.31^{0.05}_{0.05}$ 
& $0.051^{0.011}_{0.011}$ 
& $0.67^{0.05}_{0.05}$ 
& $14.1^{0.9}_{0.9}$ 
& $<0.49$ 
\vspace{2pt}
\\
\multicolumn{2}{l}{{\em strong data}+flat}
$$
& (1.00) 
& $0.97^{0.05}_{0.05}$ 
& $0.022^{0.002}_{0.002}$ 
& $0.12^{0.01}_{0.01}$ 
& $0.70^{0.04}_{0.05}$ 
& $0.30^{0.04}_{0.04}$ 
& $0.046^{0.004}_{0.004}$ 
& $0.69^{0.04}_{0.04}$ 
& $13.7^{0.2}_{0.2}$ 
& $<0.32$ 
\vspace{2pt}
\\
\multicolumn{2}{l}{{\em strong data}+flat+LSS($low$--$\sigma_8$)}
$$
& (1.00) 
& $0.97^{0.05}_{0.05}$ 
& $0.022^{0.002}_{0.002}$ 
& $0.12^{0.01}_{0.01}$ 
& $0.71^{0.05}_{0.04}$ 
& $0.28^{0.05}_{0.05}$ 
& $0.045^{0.004}_{0.004}$ 
& $0.70^{0.04}_{0.04}$ 
& $13.7^{0.2}_{0.2}$ 
& $<0.31$ 
\\
\enddata
\tablecomments{\small
Parameter estimates and errors for several prior combinations
with and without \acbar.  Errors are quoted at 1--$\sigma$ (16\% and 84\%
points of the integral of the likelihood), except for $\tau_C$ where
the 95\% upper-limit is given.  The various priors are described in 
the text.  The top block lists results found with and without the
inclusion of \acbarsp data, which shows the small improvements found
upon adding \acbarsp to the mix.  The bottom block shows the effect
of applying stronger priors on the \acbar+Others dataset, which 
naturally leads to dramatic improvements on the parameter estimates.
The difference between the LSS and LSS(low-$\sigma_8$) priors (discussed in
the text) does lead to several slight shifts, smaller
than the 1--$\sigma$ errors. }
\label{estimatetab}
\end{deluxetable}

\begin{deluxetable}{ccrrrrrrr}
\tabletypesize{\scriptsize}
\tablecaption{Eigenmodes}
\tablewidth{0pt}
\tablehead{
\colhead{Eigenmode} &
\colhead{Error} &
\colhead{$\tau_C$} &
\colhead{ $\frac{\Delta \omega_{cdm}}{\omega_{cdm}}$ } &
\colhead{ $\frac{\Delta \omega_b}{\omega_b}$ } &
\colhead{$\Omega_\Lambda$} &
\colhead{$\Omega_k$} &
\colhead{$n_s$} &
\colhead{$C_{10}$}
\vspace{2pt}
}
\startdata
\multicolumn{8}{l}{``Others"=Archeops+Boomerang+CBI+DASI+DMR+Maxima+VSA} \\
1 &  0.012 & -0.233 &  0.082 & -0.084 &  0.111 & -0.724 &  0.604 &  0.174 \\ 
2 &  0.017 & -0.210 & -0.223 & -0.007 & -0.238 &  0.607 &  0.676 &  0.155 \\ 
3 &  0.046 &  0.432 &  0.236 & -0.751 &  0.146 &  0.111 &  0.231 & -0.327 \\ 
4 &  0.091 &  0.107 &  0.700 &  0.521 &  0.315 &  0.196 &  0.244 & -0.171 \\ 
5 &  0.126 &  0.421 &  0.007 &  0.260 & -0.750 & -0.231 &  0.157 & -0.340 \\ 
\\ 
\multicolumn{8}{l}{``\acbar+Others"} \\
1 &  0.010 & -0.140 &  0.133 & -0.051 &  0.194 & -0.889 &  0.344 &  0.123 \\ 
2 &  0.015 &  -0.264 &  -0.155 &  -0.026 &  -0.238 & 0.320 & 0.843 & 0.189 \\ 
3 &  0.043 &  0.489 &  0.339 & -0.673 &  0.070 &  0.081 &  0.259 & -0.338 \\ 
4 &  0.063 &  0.368 &  0.279 &  0.362 & -0.780 & -0.198 &  0.050 & -0.078 \\ 
5 &  0.088 & 0.012 & 0.630 & 0.541 & 0.451 & 0.223 & 0.206 &  -0.123 \\ 
\enddata
\tablecomments{\small
Eigenmodes and the uncertainties on their determination, 
for the ``Others" analysis (top set) and the 
``\acbar+Others" analysis (bottom set).  Only the top five (of seven) 
are listed.  The first column labels
the eigenmodes in rank order of uncertainty;  these uncertainties,
and the eigenvectors chosen, are derived from 
weak-$h$ prior ensemble averages over the database as described
in the text.
The second column lists the uncertainty on each eigenmode, while
columns 3-9 list the coefficients of the eigenmode rotation matrix
$R_{ak}$, applied to the basis set of parameters labelled 
at the top of the columns.  
There is significant improvement,
especially for the fifth eigenmode, upon adding \acbarsp
to the dataset; note that it becomes the fourth eigenmode of
``\acbar+Others", and shows a factor of two improvement.}
\label{tab:eigenmodes}
\end{deluxetable}

\begin{deluxetable}{lll}
\tabletypesize{\scriptsize}
\tablecaption{$\sigma_8^{\rm SZ}$ Estimates}
\tablewidth{0pt}
\tablehead{
\colhead{Data}
& \colhead{SPH}
& \colhead{Analytic}
}
\startdata
ACBAR
& $1.04^{0.17}_{0.48}$ 
& $1.12^{0.18}_{0.55}$ 
\vspace{3pt}
\\
ACBAR+CBI+BIMA
& $1.04^{0.11}_{0.21}$ 
& $0.98^{0.12}_{0.21}$
\vspace{3pt}
\\
ACBAR+CBI+BIMA$^\dagger$
& $1.04^{0.12}_{0.42}$
& $0.96^{0.15}_{0.41}$
\\
\enddata
\tablecomments{\small
The results for the phenomenological fits to a SZE component. The
table shows $\sigma_8^{\rm SZ}$ estimates with statistical 
2--$\sigma$ errors, for both SPH and analytical models.  
Non--Gaussian corrections are included.\\
$^\dagger$ No lognormal transformation applied to the BIMA band power.}
\label{tabsz}
\end{deluxetable}

\end{document}